\documentclass[11pt]{article}
\usepackage{graphicx}

\newcommand{\BABARPubYear}    {04}

\newcommand{\BABARConfNumber} {035}
\newcommand{\SLACPubNumber} {10638}
\newcommand{\LANLNumber} {0408081}

\input pubboard/babarsym

\newcommand{\etal}{{\it et al.}}

\def\Btopipi   {\ensuremath{\B \to \pi\pi}\xspace}
\def\Btopipiz   {\ensuremath{\Bpm \to \pipm\piz}\xspace}
\def\Btokpi   {\ensuremath{\B \to K\pi}\xspace}
\def\Btokpiz   {\ensuremath{\Bpm \to \Kpm\piz}\xspace}
\def\Btohpiz   {\ensuremath{\Bpm \to \hpm\piz}\xspace}
\def\Bztopizpiz   {\ensuremath{\Bz \to \piz\piz}\xspace}
\def\BzBzbartopizpiz   {\ensuremath{\Bz (\Bzb) \to \piz\piz}\xspace}
\def\Bzbartopizpiz   {\ensuremath{\Bzb \to \piz\piz}\xspace}

\def\Btorhopi   {\ensuremath{\B \to \rho\pi}\xspace}
\def\Btorhozpi   {\ensuremath{\Bpm \to \rho^0\pi^\pm}\xspace}
\def\Btorhoppim   {\ensuremath{\Bz \to \rho^\pm\pi^\mp}\xspace}
\def\Btorhok   {\ensuremath{\Bz \to \rho^\pm K^\mp}\xspace}
\def\Btorhopiz   {\ensuremath{\Bpm \to \rho^{\pm}\piz}\xspace}
\def\Bptorhoppiz   {\ensuremath{\B^+ \to \rho^+\piz}\xspace}
\def\Bmtorhompiz   {\ensuremath{\B^- \to \rho^-\piz}\xspace}
\def\Btokstppiz   {\ensuremath{\Bpm \to {\Kstar}^\pm\piz}\xspace}
\def\Bztokstzpiz   {\ensuremath{\Bz \to {\Kstar}^0\piz}\xspace}
\def\Btag {\ensuremath{B_{\rm tag}}}

\def\hpm    {\ensuremath{h^{\pm}}\xspace} 

\def\acp {\ensuremath{{\cal A}}\xspace}
\def\acpcp {\ensuremath{{\cal A}_{\CP}}\xspace}

\def\alphaeff {\ensuremath{\alpha_{\rm eff}}\xspace}

\def\de {\ensuremath{\Delta E}\xspace}
\def\cossph   {\ensuremath{|\cos{\theta_{\scriptscriptstyle S}}|\;}\xspace}
\def\fish    {\ensuremath{\cal F}\xspace}
\def\mz        {\mbox{$m_{0}$}\xspace}

\long\def\inst#1{\par\nobreak\kern 4pt\nobreak
    {\it #1}\par\vskip 10pt plus 3pt minus 3pt}

\begin{document}
{\pagestyle{empty}

\begin{flushright}
\babar-CONF-\BABARPubYear/\BABARConfNumber \\
SLAC-PUB-\SLACPubNumber \\
hep-ex/\LANLNumber \\
August 2004 \\
\end{flushright}

\par\vskip 5cm

\begin{center}
\Large \bf Study of  \BzBzbartopizpiz, \Btopipiz and \Btokpiz decays
\end{center}
\bigskip

\begin{center}
\large The \babar\ Collaboration\\
\mbox{ }\\
\today
\end{center}
\bigskip \bigskip

\begin{center}
\large \bf Abstract
\end{center}

We present updated measurements of the
branching fractions for the modes \Bztopizpiz, \Btopipiz, and \Btokpiz.
We also measure the time-integrated asymmetry $C_{\piz\piz}$ and the charge
asymmetries  $\acp_{\pipm\piz}$ and $\acp_{K^\pm\piz}$.
Based on a sample of approximately 227 million 
\BB\ pairs collected by the \babar\ detector at the \pep2\
asymmetric-energy $B$ Factory at SLAC,
we measure 
\begin{eqnarray*}
\BR(\Bztopizpiz) = (1.17 \pm 0.32 \pm 0.10) \times 10^{-6} \ , \
C_{\piz\piz}= -0.12 \pm 0.56 \pm 0.06 \ ,
\end{eqnarray*}
where the first errors are statistical and the second are systematic.  The
\Bztopizpiz signal has a significance of $4.9~\sigma$ including
systematic uncertainties. 
We also measure 
\begin{eqnarray*}
{\BR}(\Btopipiz)   = (5.8 \pm 0.6 \pm 0.4) \times 10^{-6} \  , \ \acp_{\pipm\piz}   = -0.01\pm 0.10\pm 0.02 \ ,
\\
{\BR}(\Btokpiz)   = (12.0 \pm 0.7 \pm 0.6) \times 10^{-6} \  , \ \acp_{\Kpm\piz}   = 0.06 \pm 0.06 \pm 0.01 \ .
\end{eqnarray*}
Using related \babar \ measurements and isospin relations we   
find an upper bound on the angle difference  $\left| \delta \right| =\left| \alpha - \alpha_{\rm eff} \right|$
of $35^{\rm o}$ at the $90\%$ C.L.

\vfill
\begin{center}

Submitted to the 32$^{\rm nd}$ International Conference on High-Energy Physics, ICHEP 04,\\
16 August---22 August 2004, Beijing, China

\end{center}

\vspace{1.0cm}
\begin{center}
{\em Stanford Linear Accelerator Center, Stanford University, 
Stanford, CA 94309} \\ \vspace{0.1cm}\hrule\vspace{0.1cm}
Work supported in part by Department of Energy contract DE-AC03-76SF00515.
\end{center}

\newpage
} 

\begin{center}
\small

The \babar\ Collaboration,
\bigskip

%
B.~Aubert,
R.~Barate,
D.~Boutigny,
F.~Couderc,
J.-M.~Gaillard,
A.~Hicheur,
Y.~Karyotakis,
J.~P.~Lees,
V.~Tisserand,
A.~Zghiche
\inst{Laboratoire de Physique des Particules, F-74941 Annecy-le-Vieux, France }
% ...

\end{center}\newpage

\section{INTRODUCTION}
\label{sec:Introduction}

The study of \B meson decays into charmless hadronic final states
is important for the understanding of \CP violation in the
\B system. In the Standard Model, \CP violation arises from a single
phase in the Cabibbo-Kobayashi-Maskawa quark-mixing matrix $V_{\rm qq^\prime}$~\cite{ckmref}.
Measurements of the time-dependent \CP-violating asymmetry in the
\Bztopipi decay mode by the \babar\
and Belle collaborations~\cite{babarsin2alpha,bellesin2alpha}  provide information on the angle
$\alpha \equiv \arg\left[-V_{\rm td}^{}V_{\rm tb}^{*}/V_{\rm ud}^{}V_{\rm ub}^{*}\right]$ of
the Unitarity Triangle. However, in contrast to the theoretically
clean determination of the angle $\beta$ in \Bz decays to charmonium
final states~\cite{babarsin2beta,bellesin2beta}, the extraction of 
$\alpha$ in \Bztopipi is complicated by the interference of tree and
penguin amplitudes with different weak phases.  The difference between
the angles $\alpha$ and \alphaeff,
where \alphaeff is
derived from the measured time-dependent \Bztopipi \CP asymmetry, 
may be evaluated using measurements
of the isospin-related decays $\Bz (\Bzb) \to \piz\piz$ and
$\Bpm\to\pipm\piz$~\cite{Isospin}. The amplitudes $A^{ij}$ for the $B\to \pi^i\pi^j$ decays
satisfy the relation 
\begin{equation}\label{eq:isospin}
A^{+0} = \frac{1}{\sqrt{2}}A^{+-} + A^{00},
\end{equation}
and a similar relation for the conjugate amplitudes $\overline{A}^{+-}$, $\overline{A}^{00}$
and $\overline{A}^{+0}=A^{-0}$. 

For \Bztopizpiz, the \CP-related quantity that can be accessed experimentally is the time-integrated 
asymmetry $C_{\piz\piz}$, defined as
\begin{equation}
C_{\piz\piz} = \frac{ 1 - \left|\lambda_{00}\right|^2}{1+\left|\lambda_{00}\right|^2},
\end{equation}
where $\left| \lambda_{00}\right| = \left|\overline{A}_{00}/A_{00} \right|$
is the modulus of the ratio of \Bztopizpiz  and
\Bzbartopizpiz
decay amplitudes. $C_{\piz\piz}$ may deviate from zero  if the tree
and penguin amplitudes are of comparable size and have different weak and strong phases.
For \Bpm modes, the \CP-violating charge asymmetry is defined as
\begin{equation}
\acpcp =  \frac{|\overline{A}|^2 - |A|^2 }{|\overline{A}|^2 +  |A|^2 },
\end{equation}
where $A$ ($\overline{A}$) is the \Bp (\Bm) decay amplitude.
In the Standard Model,  the
decay \Btopipiz is governed by a pure tree amplitude; as a result no charge
asymmetry is expected.
Both the rate and asymmetry of the decay \Btokpiz may be used to 
extract useful constraints on penguin contributions to the \Btokpi amplitudes.

In this paper, we report
a measurement of the
time-integrated \CP asymmetry in the \Bztopizpiz decay, and an updated measurement of the
\Bztopizpiz branching fraction~\cite{babarpizpiz}. We also update the \Btopipiz and \Btokpiz branching fractions
and charge asymmetries~\cite{babarhpiz}.
This study is  based on $(226.6 \pm 2.5) \times 10^{6}$ $\FourS\to\BB$ decays 
(on-resonance data), collected with the \babar\ detector. We also use
approximately $16\invfb$ of data recorded $40\mev$ below
the \BB\ pair production threshold (off-resonance data).

\section{THE \babar\ DETECTOR AND PARTICLE SELECTION }
\label{sec:babar}

\babar\ is a solenoidal detector optimized for the asymmetric-energy
beams at \pep2 and is described in detail in Ref.~\cite{babarnim}.
Charged particle (track) momenta are measured with a 5-layer
double-sided silicon vertex tracker and a 40-layer drift chamber
inside a 1.5-T superconducting solenoidal magnet.  Neutral cluster
(photon) positions and energies are measured with an electromagnetic
calorimeter (EMC) consisting of 6580 CsI(Tl) crystals.  The photon
energy resolution is $\sigma_{E}/E = \left\{2.3 / E(\gev)^{1/4}
\oplus 1.9 \right\}
\%$, and the angular resolution from the interaction point is
$\sigma_{\theta} = 3.9^{\rm o}/\sqrt{E(\gev)}$. 
The photon energy
scale is determined using symmetric $\piz\to\gamma\gamma$ decays.
Charged hadrons are identified with a detector of internally reflected
Cherenkov light (DIRC) and  using ionization measurements in the tracking detectors. 
An instrumented magnetic-flux return detects neutral hadrons and
identifies muons.  
High efficiency for recording \BB events in which
one \B decays with low multiplicity is achieved with a two-level
trigger with complementary tracking-based and calorimetry-based
trigger decisions.

Candidate \piz\ mesons are reconstructed as pairs of photons, spatially
separated in the EMC,  
with an invariant mass $m_{\piz}$ satisfying $110<m_{\piz}<160$ \mevcc.
The resolution is approximately 8~\mevcc for high momentum \piz's.
Photon candidates are required to be consistent with 
the expected lateral shower shape, not be matched to a track, and have a
minimum energy of 30 \mev. 
To reduce the background from false \piz candidates, the 
angle $\theta_{\gamma}$ between the photon momentum vector in the \piz rest frame and
the \piz momentum vector in the laboratory frame is required to satisfy
$|\cos{\theta_{\gamma}}| < 0.95$.

Candidate tracks are required to be within the
tracking fiducial volume, originate from the interaction point,  
consist of at least 12 DCH hits, and be associated with at least 6 Cherenkov
photons in the DIRC.

\section{ANALYSIS METHOD}
\label{sec:Analysis}

\B meson candidates are reconstructed by combining a
\piz with a pion or kaon (\hpm) or by combining two \piz mesons.
Two kinematic variables, used to isolate the \Bztopizpiz and \Btohpiz signal events, take
advantage of the kinematic constraints of \B mesons produced at the
\FourS. The first is the beam-energy-substituted mass $\mes = \sqrt{
  (s/2 + {\bf p}_{i}\cdot{\bf p}_{B})^{2}/E_{i}^{2}- {\bf
    p}^{2}_{B}}$, where $\sqrt{s}$ is the total \epem center-of-mass (CM)
energy,  $(E_{i},{\bf p}_{i})$ is the four-momentum of the initial
\epem system and ${\bf p}_{B}$ is the \B candidate momentum, both measured in
the laboratory frame.  The second variable is \de 
$ = E_{B} -
\sqrt{s}/2$, where $E_{B}$ is the \B candidate energy in the CM frame.
 The \de resolution for signal is approximately 80~\mev for \Bztopizpiz,
and 40~\mev for \Btohpiz.

The primary source of background is $\epem \to \qqbar \;(q =
u,d,s,c)$ events where a \piz
or \hpm from each quark randomly combines to mimic a \B decay.
The jet-like \qqbar background is suppressed by
requiring that the angle $\theta_{\scriptscriptstyle S}$ between the
sphericity axes of the \B candidate and that of the
remaining tracks and photons in the event, in the CM frame, satisfies
$\cossph < 0.7 \,(0.8)$
for \Bztopizpiz (\Btohpiz).
The other sources of background are \B to vector-pseudoscalar decays: 
\Btorhopiz for the \Bztopizpiz mode;
\Btorhoppim, \Btorhozpi, and \Btorhopiz for the \Btopipiz mode; and \Btorhok,
\Btokstppiz, and \Bztokstzpiz for the \Btokpiz mode. 
In these decays the vector meson is
polarized, so one of the pions is often produced almost at rest in the \B rest frame, and
the remaining decay products match the kinematics of a
\Bztopizpiz or \Btohpiz decay.

For the \Bztopizpiz analysis we restrict the $\mes-\de$ plane to the region with
$\mes>5.2$ \gevcc and $|\de|<0.4$ \gev. For the on-resonance sample we
define the signal region as 
the band in the plane with $|\de|<0.2$ \gev and the
sideband region as the rest of the plane with the exception of a region at
negative \de also populated with \Btorhopiz events. The
entire plane for the off-resonance data and the sideband region for the
on-resonance data are populated with 
\qqbar background events,
which are kept in the fit in order to  constrain the corresponding background
parameters.
\Btohpiz candidates are selected in the region with $\mes>5.22$ \gevcc and  $ -0.11 < \de < 0.15 \gev $. 

For \Bztopizpiz candidates, the other tracks and clusters in the event are used to determine whether the other
$B$ meson (\Btag) decayed as a $\Bz$ or $\Bzb$ (flavor tag).
We use a multivariate technique~\cite{sin2betaPRD02} to determine the flavor of
the $\Btag$ meson.  Separate neural networks are trained to identify primary leptons, kaons,
soft pions from $D^*$ decays, and high-momentum charged particles from \B\ decays.
Each event is assigned to one of several mutually exclusive tagging categories
based on the estimated mistag probability and on the source of tagging information.

The number of signal \B candidates is determined in an extended unbinned
maximum likelihood fit.  The probability ${\cal P}_i\left(\vec{x}_j;
  \vec{\alpha}_i\right)$
for a signal or background hypothesis
is the product of probability density functions (PDFs) for the
variables $\vec{x}_j$  given the set of parameters $\vec{\alpha}_i$.
The likelihood function is given by a product over the $N$ events and
the $M$ signal and background hypotheses:
\begin{equation}
{\cal L}= \exp\left(-\sum_{i=1}^M n_i\right)\,
\prod_{j=1}^N \left[\sum_{i=1}^M N_i {\cal P}_i\left(\vec{x}_j;
\vec{\alpha}_i\right)
\right]\, .
\end{equation}
For \Bztopizpiz the coefficients $N_{i}$ are given by 
$N_{i} = \frac{1}{2}(1-s_j\acp_i) n_i$, where $s_j$ refers to the sign of the flavor tag of the other \B in  the
event $j$ and is zero for untagged events. $n_i$ and $\acp_i$ are the number of events and raw asymmetry 
for \Bztopizpiz signal, \Btorhopiz background,
and continuum background components.  The averages of the \Btorhopiz branching 
fraction and asymmetry~\cite{rhopi0rho0pibabar,rhopi0belle}
are used to fix the number of \Bptorhoppiz and \Bmtorhompiz events to $32.1\pm 5.6$. The asymmetry
$\acp_{\piz\piz}$ is related to the  time-integrated \CP asymmetry $C_{\piz\piz}$ by 
the relation
\begin{equation}
\acp_{\piz\piz} = (1-2\chi)(1-2\omega)C_{\piz\piz},
\end{equation}
where $\chi = 0.186\pm 0.004$~\cite{PDG2004} is the time-integrated mixing probability, and $\omega$ is the
mis-tagging probability.
Asymmetries in mistag rates and efficiencies are taken into account.

For \Btohpiz the probability coefficients are  $N_{i} = \frac{1}{2}(1-
q_j \acp_i) n_i $, where  $q_j$ is the
charge of the track $h$ in the event $j$, and  the fit parameters $n_i$ and $\acp_i$ are the number
of events and asymmetry
for \Btopipiz and \Btokpiz
signal, continuum, and \B background  components.
The numbers of \B background events are fixed to the
expected number of events using the current world averages
for the $\rho\pi$~\cite{rhopi0rho0pibabar,rhopi0belle,rho0pirhopibelle,rho0pirhopicleo,rhopibabar} and
$\rho K$~\cite{rhopibabar} decays, which are $18\pm 4$ and $3\pm 1$
events, respectively. 
We estimate a contribution of  $1\pm 1$   {\ensuremath{\B \to {\Kstar}\piz}\xspace} events to the background
based on the upper limit on the branching fraction~\cite{babarkstpi}. 
The uncertainties on these numbers
are dominated by the uncertainty on selection efficiencies,
because of the sensitivity to the tight requirement in \de.

The variables $\vec{x}_j$ used for \Bztopizpiz are \mes, \de, and
a Fisher discriminant \fish.
The Fisher
discriminant is an optimized linear combination of $\sum_i
p_i$ and $\sum_i p_i \cos^2{\theta_i}$,
where $p_{i}$ is the momentum and $\theta_{i}$ is the angle with
respect to the thrust axis of the \B candidate, both in the CM frame,
for all tracks and neutral clusters not used to reconstruct the \B
meson.
For both the  \Bztopizpiz signal and the \Btorhopiz
background the \mes and \de variables are correlated and therefore a 
two-dimensional PDF  from a smoothed, simulated distribution is used.
For the continuum background, the \mes distribution is modeled as a threshold function~\cite{Argus}, whose
shape parameter $\xi$ is free in the fit, and
the \de distribution as  a quadratic function whose parameters are
free in the fit.  
The PDF for the \fish variable is modeled as a parametric step  
function (PSF)\cite{babarpizpiz} for all event components.  A PSF
is a binned distribution whose parameters are the
heights of each bin.
The PSF is normalized to one, so that the number of free
parameters is equal to the number of bins minus one. 
The \fish PSF has ten bins
chosen so that each bin contains 10\% of the signal sample.
For the \Bztopizpiz signal and the \Btorhopiz background the \fish PSF parameters
are correlated with the flavor tagging, and the PSF parameters are
different for each tagging category. 
Simulation is used to determine the PSF
distributions for both \Bztopizpiz and \Btorhopiz. 
A sample of $B\to D^{(*)}n\pi$ ($n=1,2,3$) events is used to verify that the simulation reproduces
the \fish distribution.  For
\qqbar background, the \fish PSF parameters are free parameters in
the fit.

The variables $\vec{x}_j$ used for \Btohpiz are \mes, \de, the
Cherenkov angle $\theta_c$ of the \hpm track, and the Fisher discriminant \fish.  
The PDF parameters for \mes, \de, $\theta_c$, and \fish for the background are
determined using the data, while the PDFs for signal are found from a
combination of simulated events and data.  
The \mes and \de distributions for \qqbar events are treated as in
the \Bztopizpiz case, with parameters allowed to vary freely in the fit.
For the signal, the \mes and \de
distributions are both modeled as a Gaussian distribution with a low-side power law tail
 whose parameters are determined from simulation.
The mean of the signal \mes and \de distributions
are determined from the fit
to the \Btohpiz sample and  their values used to tune the \piz energy scale in the 
\Bztopizpiz analysis. The mean of \de
for the \Btokpiz mode is a function of the kaon laboratory momentum,
since a pion mass hypothesis is used.
The distribution of \fish is
modeled as a bifurcated Gaussian for the signal, whose parameters are obtained from simulation,
 and as  a double Gaussian for 
the continuum background, whose parameters 
are determined in the likelihood fit. The difference of the measured and
expected values of $\theta_c$ for the pion or kaon hypothesis,
divided by the uncertainty on $\theta_c$, is modeled as a double
Gaussian function.
A control sample of kaon and pion tracks, from the decay $\Dstarp \to
\Dz \pip \, , \Dz \to \Km \pip$, is used to parameterize
$\sigma_{\theta_c}$ as a function of the charged track momentum.

\section{RESULTS AND SYSTEMATIC UNCERTAINTIES}
\label{sec:Physics}

\begin{figure}[!tbph]
\begin{center}
  \includegraphics[width=0.49\linewidth]{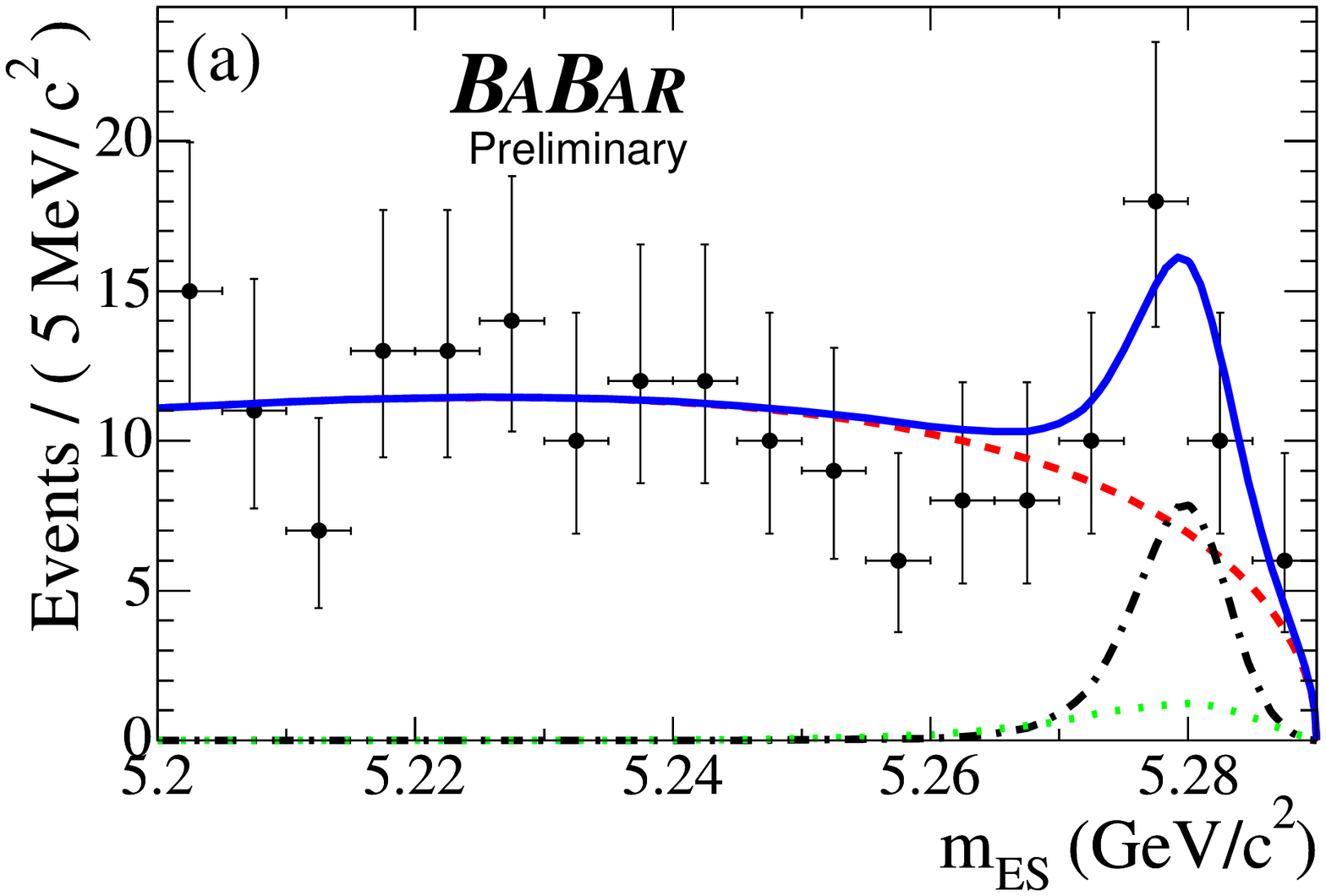}
  \includegraphics[width=0.49\linewidth]{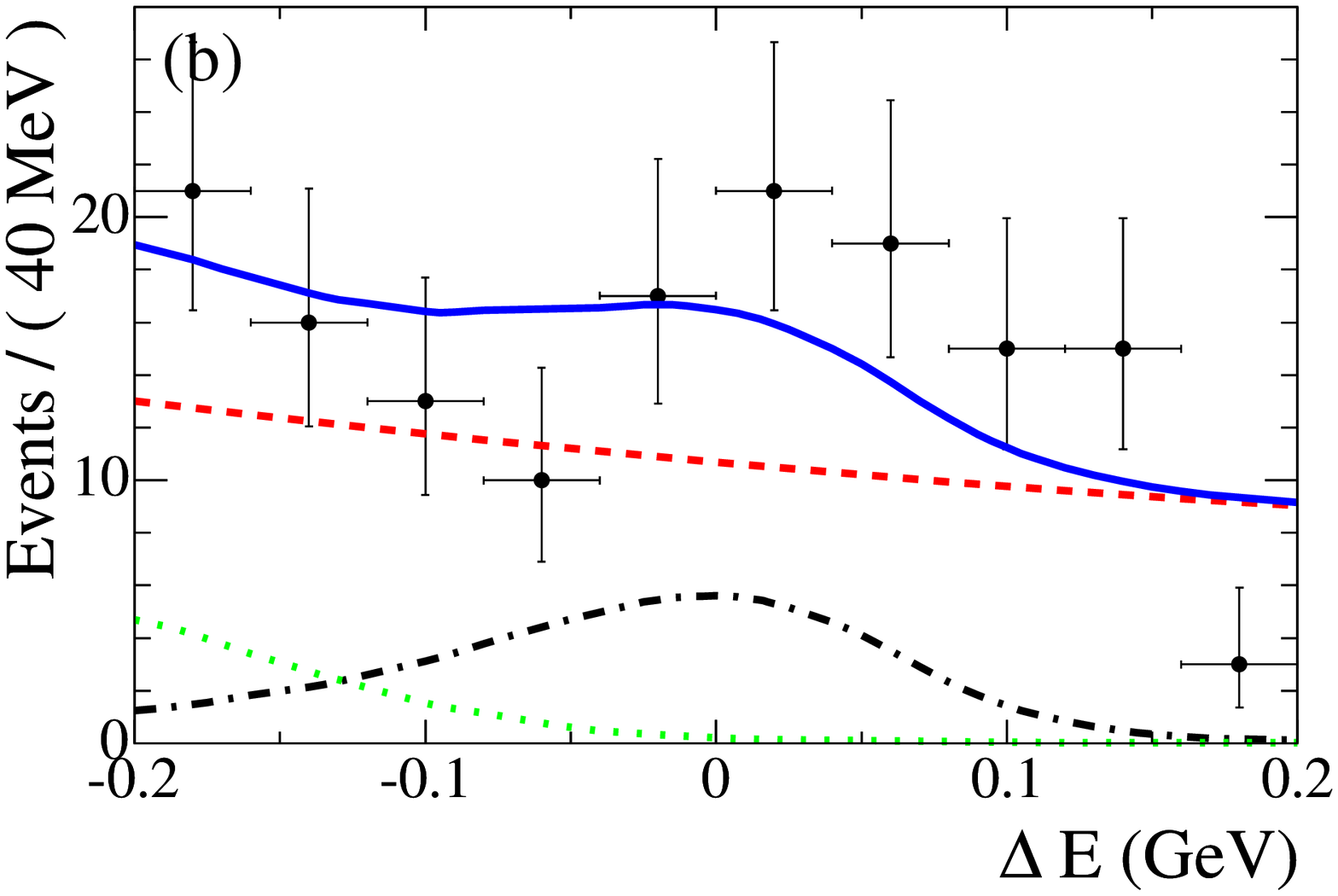}
  \includegraphics[width=0.49\linewidth]{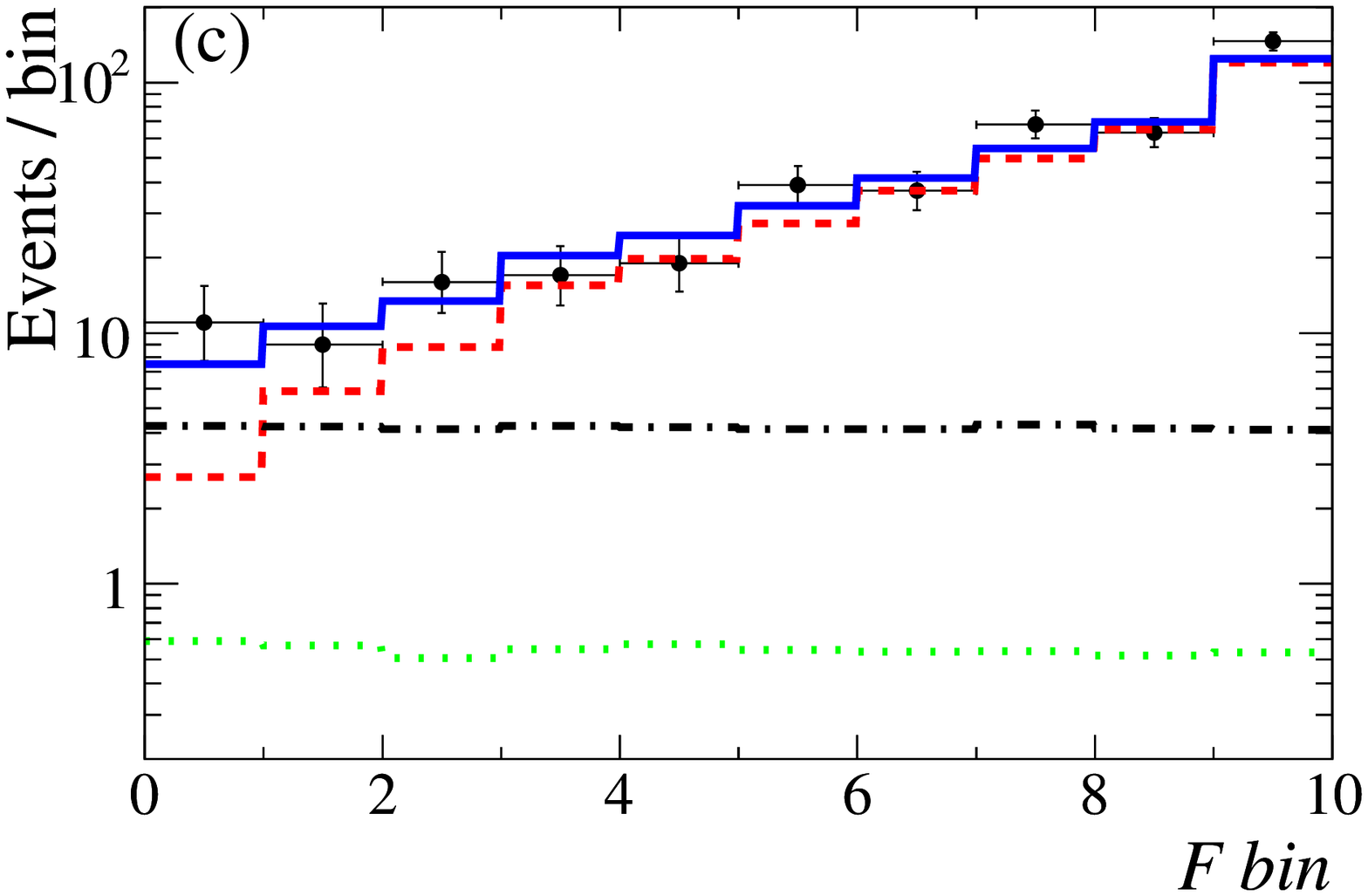}
\caption{ The  distributions of (a) \mes, (b) \de, and (c) Fisher
discriminant \fish for \Bztopizpiz
candidates in the signal data sample that satisfy an optimized
requirement on the signal probability, based on
all variables except the one being plotted.  
The projections contain 
25\%, 45\% and 68\% 
of the signal, 
14\%, 31\% and 17\%
of the $\rho\piz$
background, and 
2.2\%, 1.3\% and 4.4\% 
of the continuum background,
for \mes, \de, and \fish respectively.
The PDF projections are shown as a dashed
line for \qqbar background, a dotted line for \Btorhopiz, and a
dashed-dotted line for \Bztopizpiz signal.  The solid line shows the
sum of all PDF projections.  The PDF projections are scaled by
the expected fraction of events passing the probability-ratio requirement.}
\label{fig:pi0pi0}
\end{center}
\end{figure}

The result of the maximum likelihood fit for \Bztopizpiz is
$n(\Bztopizpiz) = 61\pm 17$ (see Table~\ref{table:summary}).
The statistical significance of the event yield
is evaluated from the square root of the change in
$-2\ln{\cal L}$ between the nominal fit and a separate fit in which
the signal yield is fixed to zero, and is found to be $5.2\sigma$
(statistical errors only).
The time-integrated asymmetry is $C_{\piz\piz} = -0.12\pm 0.56$.
Distributions of 
likelihood fit variables for \Bztopizpiz  are shown in Fig.~\ref{fig:pi0pi0}.
The data shown are for events
that have passed a probability ratio cut optimized to enhance the
signal to background fraction.
A validation is made by performing a simpler event-counting
analysis, based on the number of \Bztopizpiz candidates satisfying tighter requirements. 
The result of the event-counting analysis is $n(\Bztopizpiz) = 43\pm 26$  and
agrees well with the result from the maximum likelihood fit.  
This result is statistically consistent with our previously reported
measurement~\cite{babarpizpiz}. With changes in the analysis technique to measure
the \CP asymmetry, we now find $44\pm 13$ events in the first 123 million \BB\ events, compared
to $46\pm 13$ events found in Ref.~\cite{babarpizpiz}. The additional 104 million \BB\ events
dataset has a signal of $17\pm 11$ events. The signal rates in these two subsets agree
at the $1.3\sigma$ level. 
This result also includes an improved understanding of the \piz detection
efficiency.
Using a sample of \piz mesons from
$\tau^{\pm}\to \pipm \piz \nu_{\tau}$ decays we establish a \piz efficiency
correction of $\varepsilon_\piz = 0.99 \pm 0.03$ to our GEANT simulation,
compared to a correction of $0.88 \pm 0.08$ in
Ref.\cite{babarpizpiz}.  

\begin{figure}[!tbp]
\begin{center}
\includegraphics[width=0.49\linewidth]{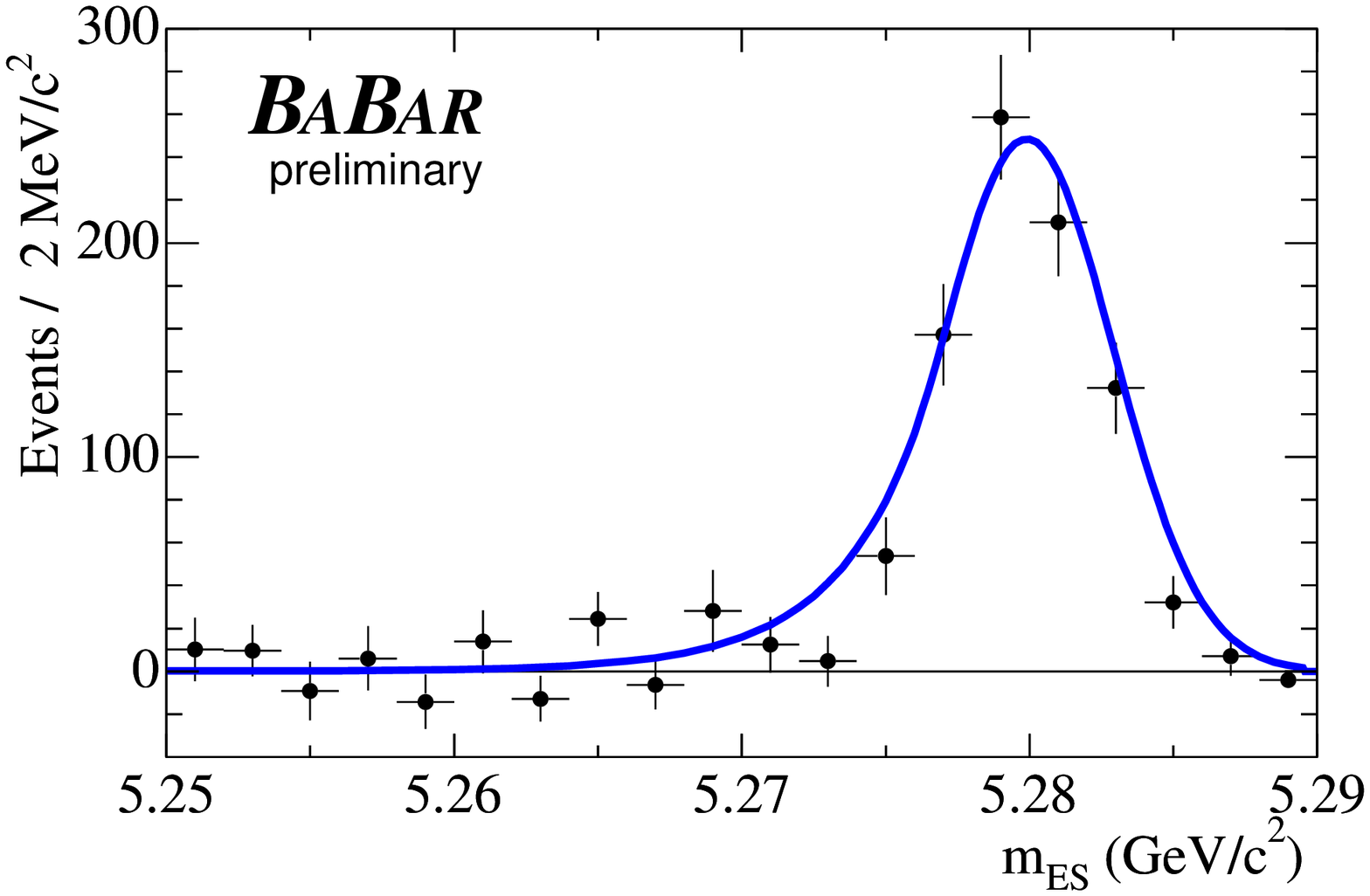}
\includegraphics[width=0.49\linewidth]{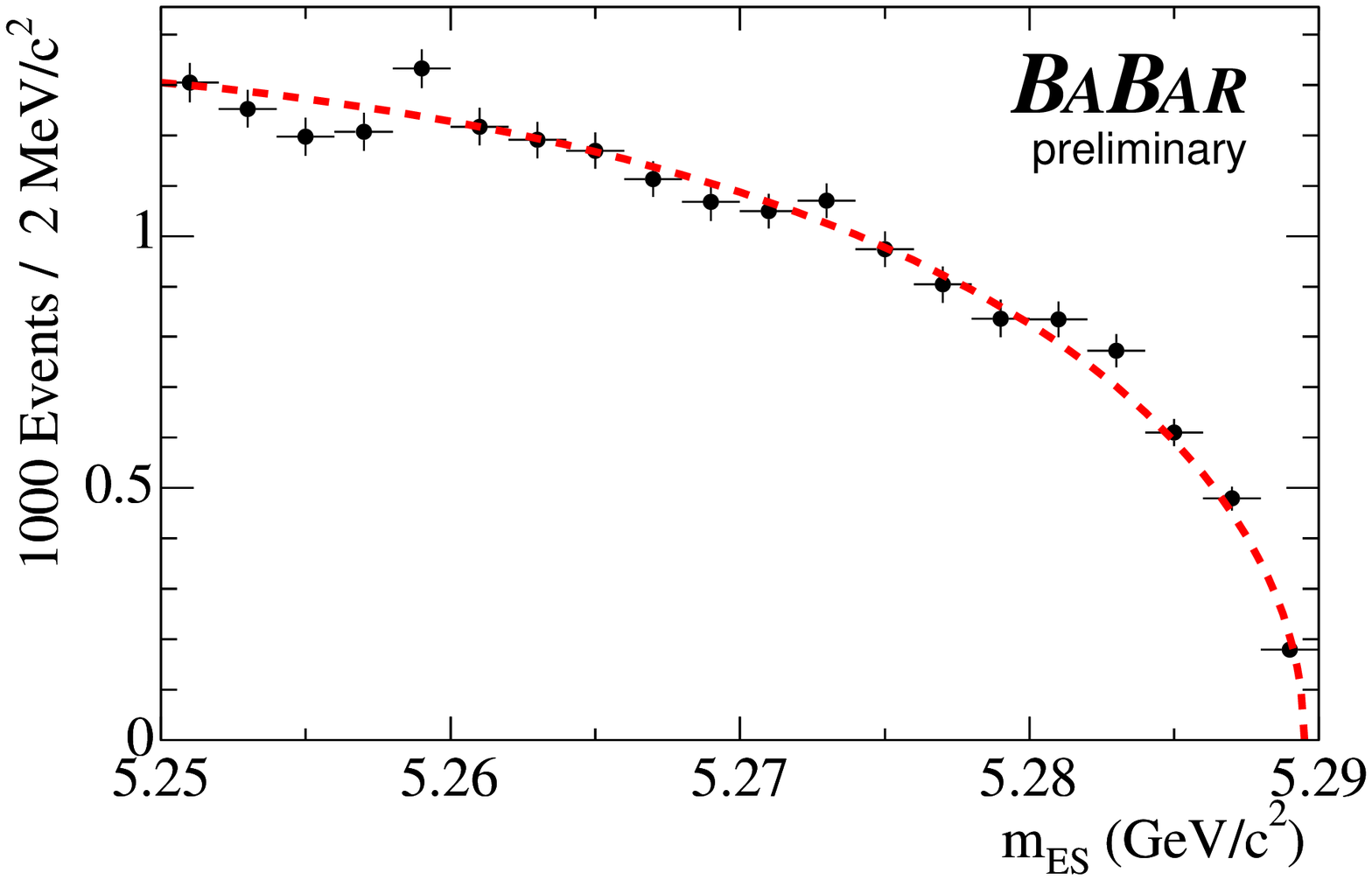}
\includegraphics[width=0.49\linewidth]{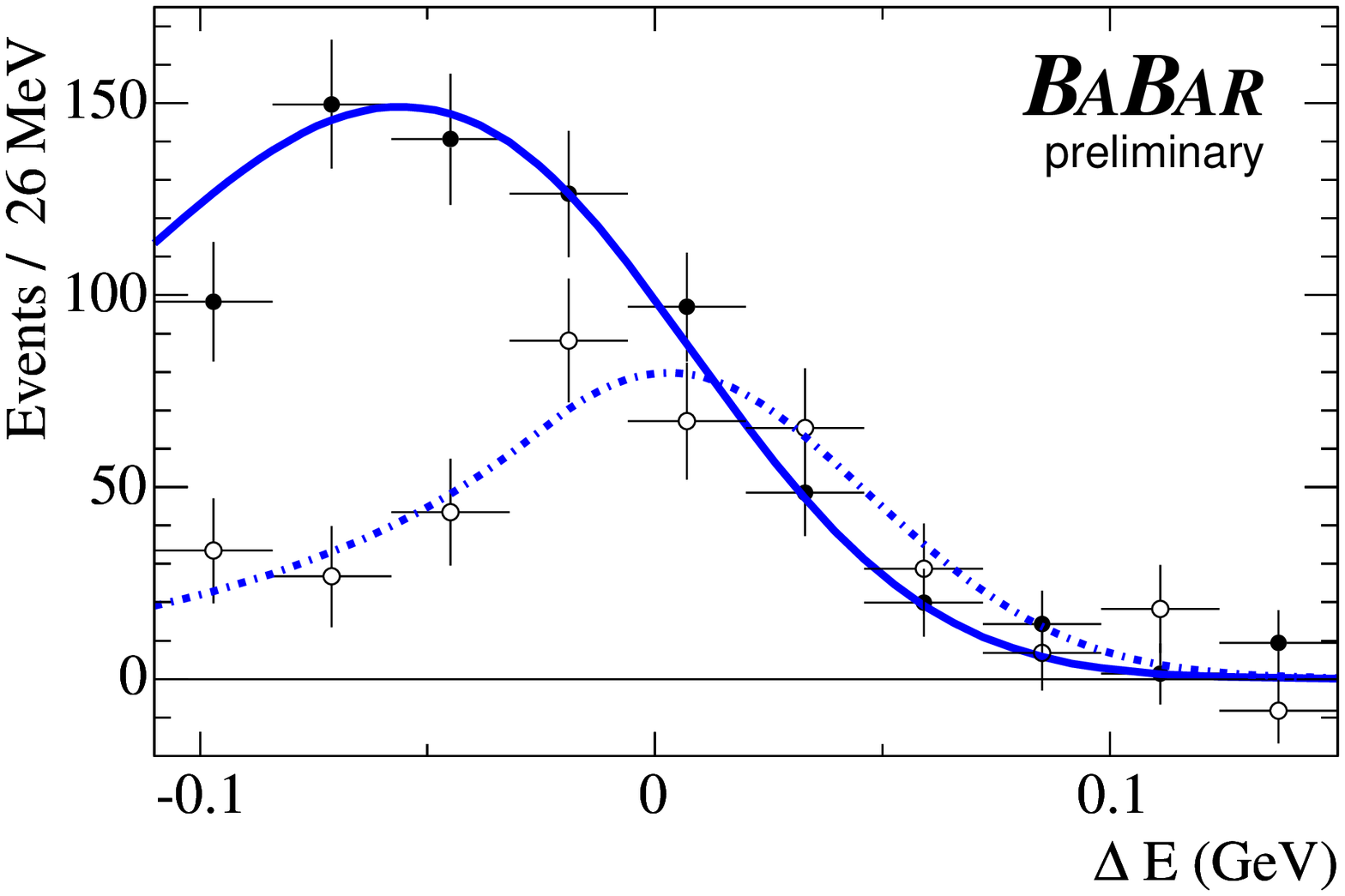}
\includegraphics[width=0.49\linewidth]{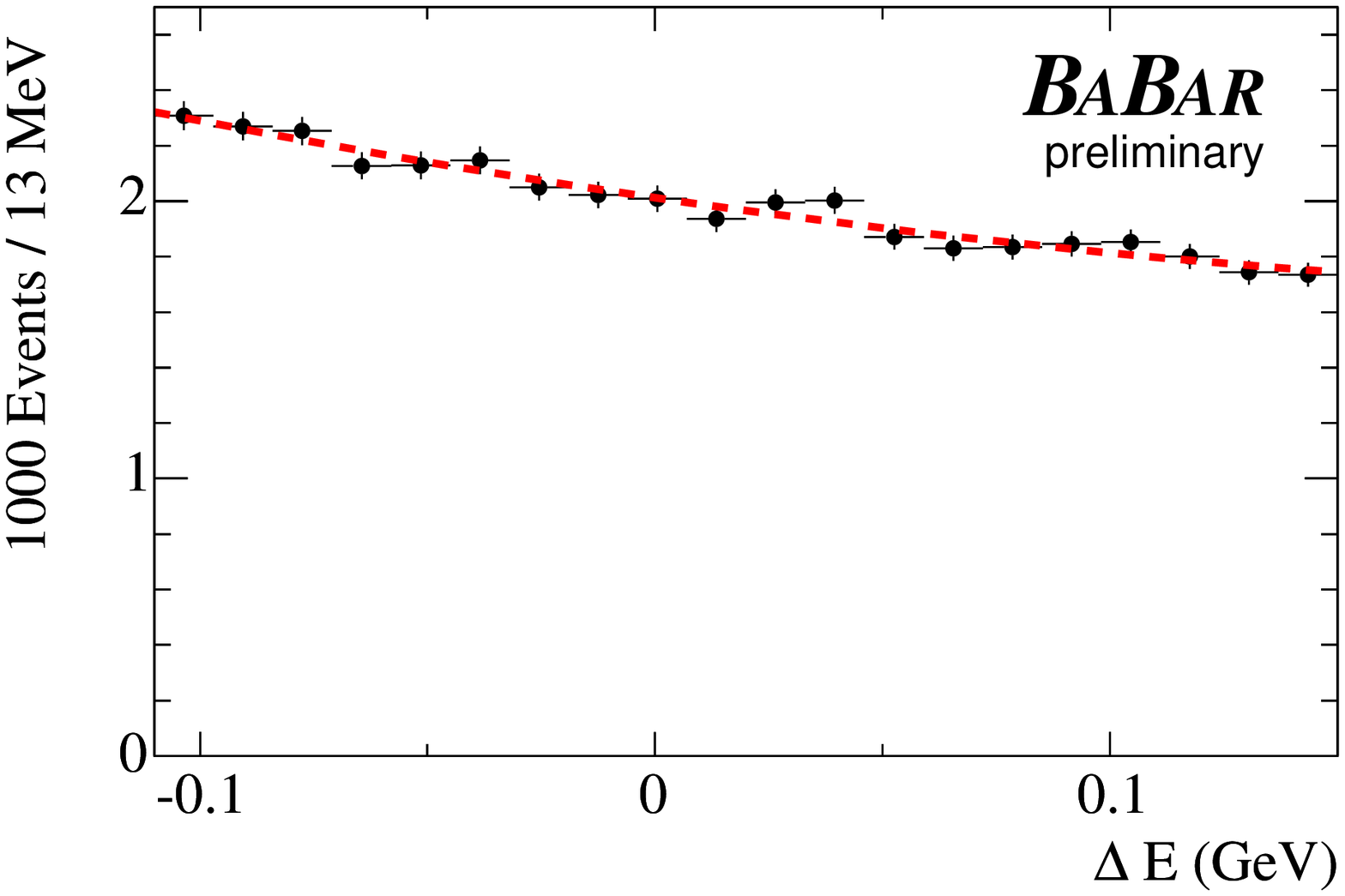}
\includegraphics[width=0.49\linewidth]{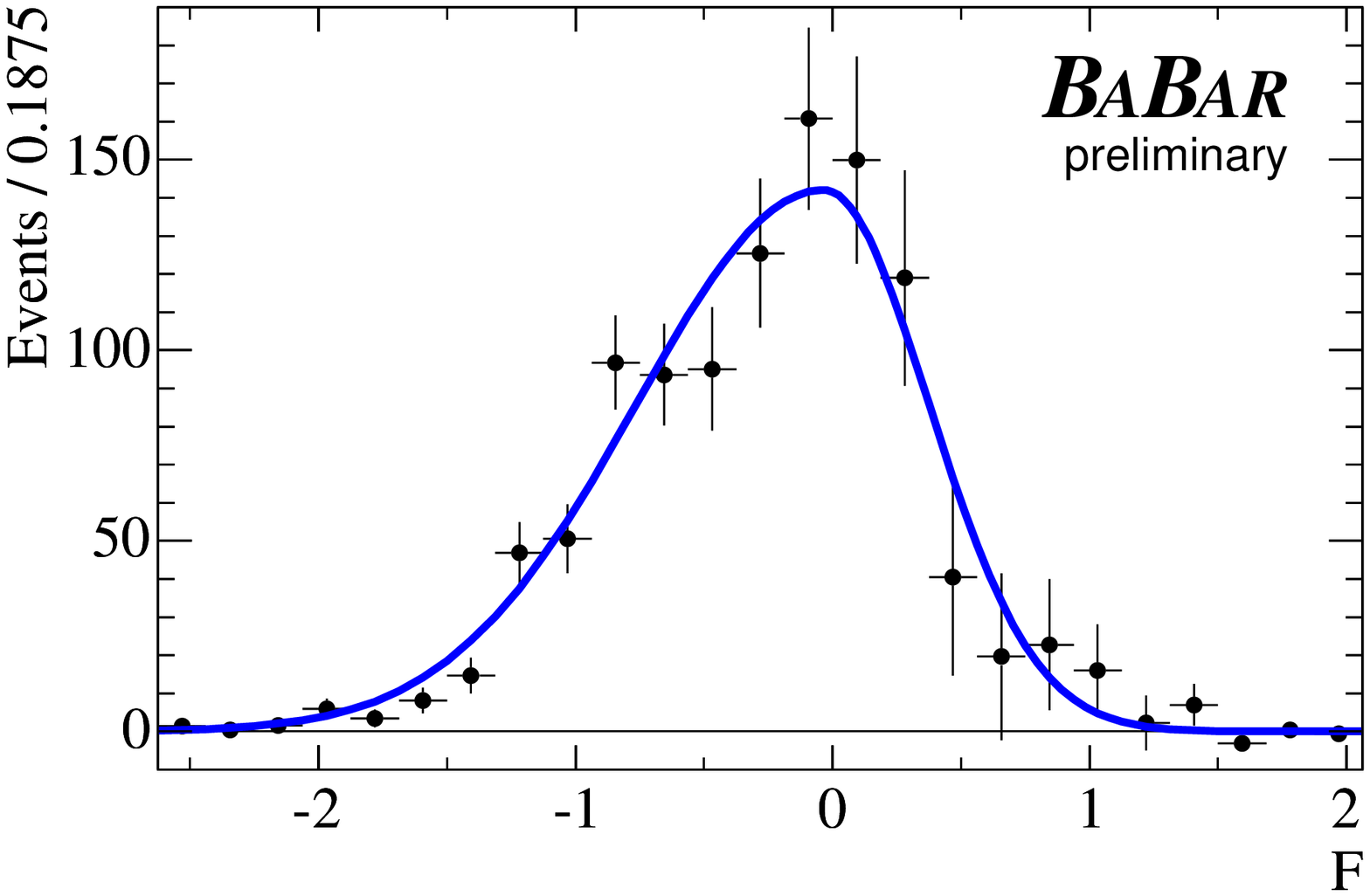}
\includegraphics[width=0.49\linewidth]{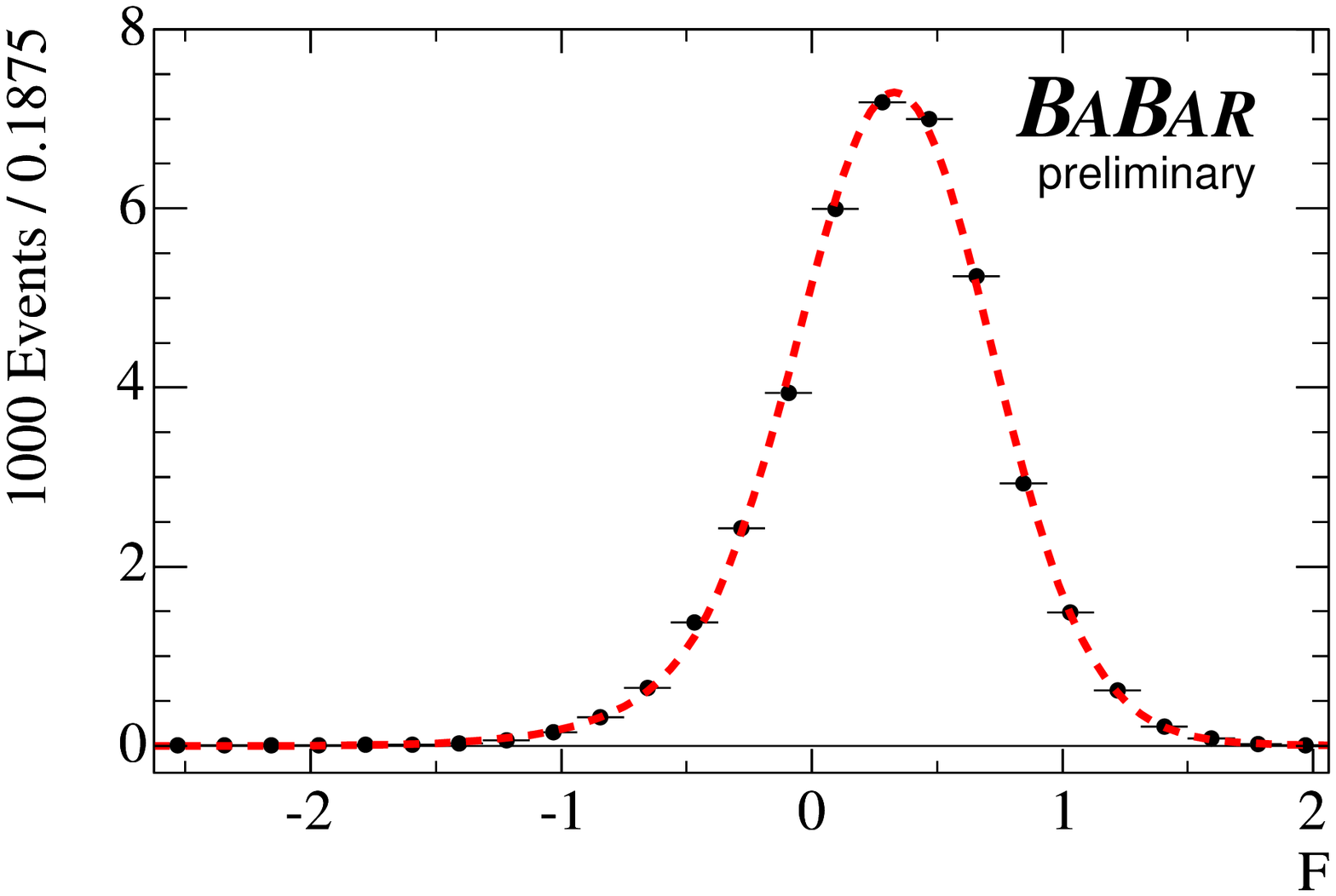}
\end{center}
\vspace{0.2cm}
\caption{
The distributions and PDF projections of \mes (top),  \de (middle) and \fish discriminant (bottom),  for
\Btopipiz and \Btokpiz candidates in the data sample. Figures on the left are for
background-subtracted signal events, and background distributions are on the right.
For \mes and \fish discriminant, the \Btopipiz and \Btokpiz distributions are combined,
while for signal \de the \Btopipiz (open circles and dotted dashed curve) and \Btokpiz (solid circles and solid curve) 
distributions are shown separately.
The background is subtracted 
using the method  described in reference \cite{splots}. The method uses all variables except
the one being plotted.
}
\label{fig:pipi0}
\end{figure}

For \Btohpiz the results are  $n(\Btopipiz) = 379\pm 41$ and
$n(\Btokpiz) = 682\pm 39$.
The charge asymmetries are  $\acp_{\pipm\piz} = -0.01\pm 0.10$
and $\acp_{\Kpm\piz} = 0.06 \pm 0.06$. 
Results are
summarized in
Table~\ref{table:summary}. Distributions of likelihood fit variables for \Btohpiz 
are shown in Fig.~\ref{fig:pipi0}.

Systematic uncertainties on the event yields and \CP asymmetries
are evaluated on data control samples, or by
varying the fixed parameters and refitting the data.
Tables~\ref{pi0pi0syst} and~\ref{hpi0syst} summarize the dominant contributions
to the systematic uncertainties for the \Bztopizpiz and \Btohpiz
branching fractions and asymmetries.
For \Bztopizpiz, the dominant systematics arise from the
uncertainty on the \de resolution and the efficiency of the \piz
reconstruction.  We reevaluate the significance of the \Bztopizpiz
signal from zero, including all systematic effects in the direction which reduces the signal, and
find a significance of $4.9\sigma$.

For both \Btopipiz and \Btokpiz, the
dominant systematic uncertainties arise from the \fish PDF parameters for
signal, selection efficiencies, and  the \de resolution.
Additional systematic uncertainties arise from
the EMC energy scale, selection efficiencies, and particle identification.
The systematic uncertainty on the charge asymmetries is 
dominated by the upper limit of the charge bias in the detector (1.0\%)~\cite{chargeasym}, since 
most selection and PDF systematics cancel in the asymmetry.

\begin{figure}[!tbp]
\begin{center}
\includegraphics[width=0.69\linewidth]{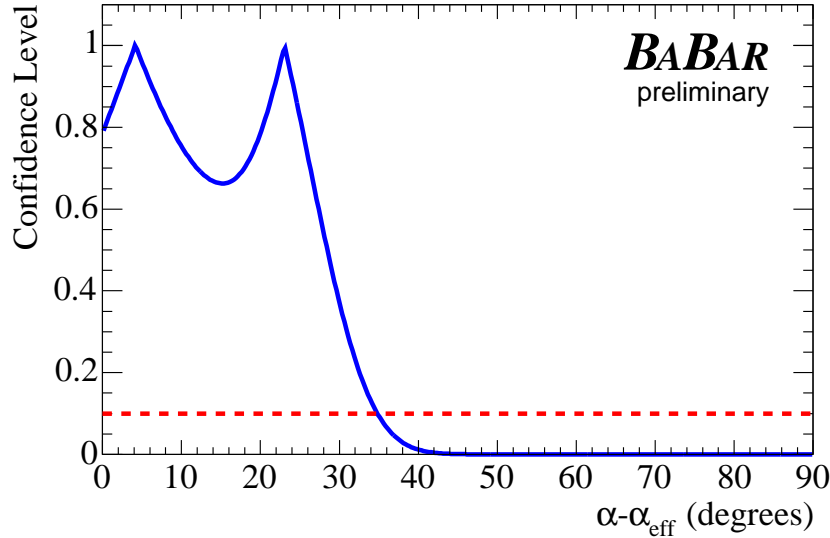}
\end{center}
\vspace{0.2cm}
\caption{ 
Constraints on the angle $\delta = \alpha-\alpha_{\rm eff}$, expressed as a Confidence Level
as a function of $|\delta|$. The constraint is evaluated using the isospin relations
(Eq. \ref{eq:isospin}) and the \babar \ measurements of
the \Btopipi branching fractions and the \CP violation parameters $C_{\piz\piz}$ and $C_{\pip\pim}$.
We find an upper bound  on $\left|\delta\right|$  of $35^{\rm o}$ at the 90\% C.L.
}
\label{fig:IsospinDalpha}
\end{figure}

We use the isospin relations, Eq.~\ref{eq:isospin}, to 
extract information on the angle difference $\delta = \alpha-\alpha_{\rm eff}~\cite{Isospin}$,
in conjunction with the \babar \ measurements of $C_{\pi^+\pi^-} = (-0.09\pm  0.15\pm 0.04)$~\cite{babarsin2alpha},
the branching fraction $\BR(\Bztopipi) = (4.7\pm 0.6\pm 0.2)\times 10^{-6}$~\cite{babarpipi}, 
the \Bztopizpiz and \Btopipiz decay rates and
the $C_{\piz\piz}$ asymmetry described in this paper.  We scan over all values of $\left| \delta \right|$
and calculate a $\chi^2$ for the five amplitudes ($A^{+-}$, $\overline{A}^{+-}$, $A^{00}$, $\overline{A}^{00}$ and
$A^{+0}$), given these five measurements 
and the two constraints (Eq.~\ref{eq:isospin} and its conjugate) for each value of
$\left| \delta \right|$. The $\chi^2$ is converted into a confidence level, as shown in Fig.~\ref{fig:IsospinDalpha},
from which we derive an upper bound on  $\left|\delta\right|$  of $35^{\rm o}$ at the 90\% C.L.

\section{CONCLUSION}
\label{sec:conclusion}

We observe $61 \pm 17 \pm 5$
\Bztopizpiz events with a significance of $4.9$~$\sigma$
including systematic uncertainties.
This corresponds to a branching fraction $\BR(\Bztopizpiz) = ( 1.17 \pm 0.32 \pm
0.10 )\times 10^{-6}$, where the first error is statistical and the
second is systematic. We measure  the time-integrated asymmetry
$C_{\piz\piz}= -0.12\pm 0.56\pm 0.06$. 
We report branching fractions  $\BR(\Btopipiz) = (5.8\pm 0.6\pm 0.4)\times
10^{-6}$ and
$\BR(\Btokpiz) = (12.0\pm 0.7 \pm 0.6)\times 10^{-6}$.
The charge asymmetries are $\acp_{\pipm\piz} =  -0.01\pm 0.10\pm 0.02$
and $\acp_{\Kpm\piz} = 0.06\pm 0.06\pm 0.01$;
we find no evidence for \CP violation.
These results are consistent with
our previous results for these
decays~\cite{babarpizpiz,babarhpiz}. We find an upper  bound on the
angle difference  $\delta = \alpha-\alpha_{\rm eff}$ of $\left|\delta\right|<35^{\rm o}$ at the 90\% C.L.

\section{ACKNOWLEDGMENTS}
\label{sec:Acknowledgments}
We are grateful for the 
extraordinary contributions of our \pep2\ colleagues in
achieving the excellent luminosity and machine conditions
that have made this work possible.
The success of this project also relies critically on the 
expertise and dedication of the computing organizations that 
support \babar.
The collaborating institutions wish to thank 
SLAC for its support and the kind hospitality extended to them. 
This work is supported by the
US Department of Energy
and National Science Foundation, the
Natural Sciences and Engineering Research Council (Canada),
Institute of High Energy Physics (China), the
Commissariat \`a l'Energie Atomique and
Institut National de Physique Nucl\'eaire et de Physique des Particules
(France), the
Bundesministerium f\"ur Bildung und Forschung and
Deutsche Forschungsgemeinschaft
(Germany), the
Istituto Nazionale di Fisica Nucleare (Italy),
the Foundation for Fundamental Research on Matter (The Netherlands),
the Research Council of Norway, the
Ministry of Science and Technology of the Russian Federation, and the
Particle Physics and Astronomy Research Council (United Kingdom). 
Individuals have received support from 
CONACyT (Mexico),
the A. P. Sloan Foundation, 
the Research Corporation,
and the Alexander von Humboldt Foundation.

\newpage
\begin{table*}[!htb]
\begin{center}
\caption{ The results for the modes \Bztopizpiz and \Btohpiz are summarized.  For each mode, the
sample size $N$, number of signal events $N_S$, total detection efficiency $\varepsilon$,
branching fraction \BR, asymmetry ${\cal A}$ or $C_{\piz\piz}$, and $90\%$ 
confidence interval for the asymmetry are shown.
For $C_{\piz\piz}$ the confidence interval is normalized to the physical region $[-1, 1]$.
Errors are statistical and systematic respectively, with the exception of $\varepsilon$ whose
error is purely systematic.}
\label{table:summary}
\begin{tabular}{l|cccccc}
Mode        & $N$    & $N_{S}$    & $\varepsilon$ (\%) & \BR($10^{-6}$)       & Asymmetry      & $(90\% {\rm C.L.})$  
\\
\hline 
\Bztopizpiz &   $8153$  &   $61\pm 17$ & $23.5\pm 1.4$   &  $1.17\pm 0.32\pm 0.10$ &  $-0.12\pm 0.56\pm 0.06$ &  $[-0.88, 0.64]$
\\
\Btopipiz   &  $29950$  & $379\pm 41$ &  $28.7\pm 1.1$ &  $5.8\pm 0.6\pm 0.4$ & $-0.01\pm 0.10\pm 0.02$ &  $[-0.19 , 0.21]$ 
\\
\Btokpiz    &  $13165$  &  $682\pm 39$ & $25.0\pm 1.0$    & $12.0\pm 0.7\pm 0.6$ & $ \ \ 0.06\pm 0.06\pm 0.01$ &  $[-0.06 , 0.18]$
\\
\end{tabular}
\end{center}
\end{table*}
\begin{table*}[!htb]
\begin{center}
\caption{ Systematic uncertainties in the determination of the
\Bztopizpiz branching fraction (left), and the $C_{\piz\piz}$ asymmetry (right).
In order of decreasing importance the branching fraction is affected by the
uncertainties in the efficiency (3\%
for each \piz), in the difference between simulation and data of the \de resolution,
in the branching fraction \Btorhopiz, and in the number of produced
$B\overline{B}$ pairs. The systematic uncertainty on $C_{\piz\piz}$ is dominated by the uncertainty
on the \B background asymmetry.}
\label{pi0pi0syst}
\begin{tabular}{l|cc}
\\
Source  &  $\Delta \BR (\piz\piz)$      &  
\\
\hline
\\
\piz efficiency & $+6\%$ & $-6\%$ \\
\de resolution  & $+7.2\%$ & $-1.0\%$ \\
\BR (\Btorhopi) & $+3.4\%$ & $-3.1\%$ \\
mean of \de and \mes & $+1.5\%$ & $-1.6\%$ \\
luminosity & $+1.1\%$ & $-1.1\%$ \\
\end{tabular}
\begin{tabular}{l|c}
\\
Source  &  $\Delta (C_{\piz\piz})$
\\
\hline
\\
\B background asymmetry &       $\pm 0.05$
\\
tagging efficiency      &       $\pm 0.03$
\\
simulation              &       $\pm 0.02$
\\
\BR (\Btorhopi)         &       $\pm 0.01$
\\
 & 
\\
\end{tabular}
\end{center}
\end{table*}
\begin{table}[!htb]
\begin{center}
\caption{ Dominant systematic uncertainties for \Btopipiz and \Btokpiz, listed in
order of decreasing importance as percent changes in the branching
fractions \BR \ (left), and absolute changes in the asymmetries $\acp_{\pipm\piz}$, $\acp_{K^\pm\piz}$ (right). 
}
\label{hpi0syst}
\begin{tabular}{l|cccc}
\\
Source  & $\Delta \BR (\pi^\pm\piz)$ & $\Delta \BR (K^\pm\piz)$ 
\\
\hline
\\
\fish, \de simulation  &   $\pm 4.2\%$    &    $\pm 3.3\%$
\\
\piz efficiency & $\pm 3.1\%$  &  $\pm 3.1\%$   
\\
\B background   &      $\pm 2.2\%$        &    $\pm 0.2\%$     
\\
luminosity      &       $\pm 1.1\%$     &       $\pm 1.1\%$ 
\\
$h^\pm$  identification &   $\pm 0.5\%$   &   $\pm 0.6\%$   
\\
\end{tabular}
\begin{tabular}{l|cc}
\\
Source  &   $\Delta(\acp_{\pipm\piz})$  &  $\Delta(\acp_{K^\pm\piz})$
\\
\hline
\\
detector bias &   $\pm 0.010$ &  $\pm 0.010$
\\
\fish, \de simulation              &   $\pm 0.011$  &  $\pm 0.007$
\\
\B background &   $\pm 0.008$  &  $\pm 0.001$
\\
$h^\pm$  identification      &       $\pm 0.004$     &       $\pm 0.003$
\\
&	&
\end{tabular}
\end{center}
\end{table}

\begin{thebibliography}{99}
\bibitem{Argus} The threshold function used for the \mes PDF is $\frac{\mes}{\mz} \sqrt{1 -
(\frac{\mes}{\mz})^{2}} \, {\rm exp}\left\{-\xi\left[1 - (\frac{\mes}{\mz})^{2}\right]\right\}$,
where \mz is the \mes endpoint, and $\xi$ the shape parameter.  See
 ARGUS Collaboration, H.~Albrecht \etal, \zpc{48}, 543 (1990).
\end{thebibliography}
\end{document}